# Not All That Can Be Automated Should Be Automated – Strategic Minimalism as a Disciplined and Ethically Grounded Approach to AI Adoption

**Victor Frimpong,** **ORCID:** https://orcid.org/0009-0004-9560-3927
DBA, Lecturer and Researcher, Management Department, SBS Swiss Business School, Kloten-Zurich, Switzerland

**Corresponding author: Victor Frimpong,** **v.frimpong@research.sbs.edu**
**Type of manuscript:** theoretical paper

**Abstract:** *As organizations accelerate the adoption and use of artificial intelligence, a common misconception arises that equates automation with progress and strategic necessity. This paper argues that not all that can be automated should be automated. It introduces Strategic Minimalism, a disciplined and ethically grounded approach to AI adoption that values purpose, proportionality, and prudence over speed and scale. Drawing on bounded rationality, virtue ethics, and frugal innovation, the study reframes technological restraint as a form of higher intelligence and responsible governance. Methodologically, the framework emerges from a structured conceptual synthesis and abductive reasoning process: the three traditions are integrated into two core dimensions (automation intensity and retained human judgment), whose intersection yields four strategic quadrants (Virtuous Minimalism, Balanced Synergy, Automation Excess, and Performative Minimalism). The framework is then conceptually stress-tested against rival explanations and boundary conditions to ensure internal coherence and analytic validity. By classifying automation into necessary, excessive, and performative types, the paper advances Responsible AI scholarship by identifying restraint as a strategic virtue that strengthens accountability, human oversight, and organizational resilience. Ultimately, it contends that in the age of intelligent machines, leadership wisdom lies not in automating more, but in knowing precisely when and why to stop.*

**Received:** 24 September 2025 **Accepted:** 01 December 2025 **Published:** 31 December 2025

**Funding:** There is no funding for this research.
**Publisher:** Academic Research and Publishing UG (i.G.) (Germany).
**Founder:** Academic Research and Publishing UG (i.G.) (Germany).

**Cite as:** Frimpong, V. (2025). Not All That Can Be Automated Should Be Automated – Strategic Minimalism as a Disciplined and Ethically Grounded Approach to AI Adoption. *Business Ethics and Leadership, 9*(4), 57-66. https://doi.org/10.61093/bel.9(4).57-66.2025.

## INTRODUCTION

Organizations across sectors are accelerating the adoption of Artificial Intelligence (AI) to enhance efficiency, accuracy, and decision-making. However, the assumption that more automation necessarily leads to better outcomes is increasingly being challenged.

In critical domains such as healthcare, finance, and governance, the unexamined expansion of automation can erode qualities that define sound human judgment (empathy, moral reasoning, and contextual understanding) (Vudugula et al., 2023; Büber & Seven, 2025). While AI can augment analytical capacity, it cannot replicate the ethical deliberation that complex decisions demand.

This paper argues that not all tasks that can be automated should be automated, and that organizational intelligence in the AI era depends as much on discernment as on capability. Existing scholarship on responsible and ethical AI focuses primarily on risk mitigation, transparency, and compliance. However, few frameworks treat restraint itself as a strategic and moral competency (an active principle guiding when automation should pause rather than proceed). To address this gap, the present study develops the concept of *Strategic Minimalism*, an integrative framework grounded in bounded rationality (Simon, 1957), virtue ethics (Anscombe, 1958; Moosmayer et al., 2025), and frugal innovation (Hossain, 2020).

Methodologically, the framework arises from structured conceptual synthesis and abductive reasoning. It identifies two critical dimensions (automation intensity and retained human judgment), whose intersection generates four strategic quadrants that define organizational posture toward AI use. These conceptual categories are further stress-tested through boundary analysis and rival theoretical explanations to ensure coherence and generalizability across contexts.

Strategic Minimalism thus reframes restraint as a form of higher intelligence, one that privileges proportionality, purpose, and prudence over expansion and excess. It argues that the sustainable value of AI does not lie in the volume of automation but in the wisdom of its selective application. The argument, therefore, advances beyond the conventional discourse of responsible AI by positioning restraint as a deliberate and measurable component of strategic intelligence. In doing so, it aligns technological governance with moral reasoning and contextual adequacy.

To operationalize this argument, the paper develops the Strategic Minimalism Framework, derived from an integrative synthesis of bounded rationality, virtue ethics, and frugal innovation. These three theoretical traditions jointly illuminate how organizations can calibrate automation to context, proportion, and moral purpose. The framework's structure, comprising two intersecting dimensions of automation intensity and retained human judgment, provides a conceptual map for identifying different organizational postures toward AI use. The following section explicates these theoretical foundations, showing how their integration redefines restraint as both an ethical stance and a strategic resource in the age of intelligent machines.

## LITERATURE REVIEW

Strategic Minimalism combines three theoretical traditions: bounded rationality, virtue ethics, and frugal innovation. These perspectives redefine restraint as a sign of intelligence rather than a limitation. They offer a solid foundation for viewing constraint as a productive and ethical choice in AI strategy.

### *Bounded Rationality and the Limits of Optimization*

Herbert Simon's theory of bounded rationality (1957) emphasizes that decision-makers face cognitive, informational, and time constraints. AI can help overcome these limitations by quickly processing large amounts of data and providing insights.

However, it can create a false sense of knowing everything, leading to over-reliance on AI and reducing human involvement in decision-making. Westphal et al. (2025) also found that AI enhances task performance and satisfaction in delegation scenarios, highlighting the importance of human self-efficacy in using these technologies effectively.

Strategic Minimalism values adequate intelligence applied in context over maximum intelligence applied without discretion. Mândricel (2025) emphasizes that hybrid governance models integrating AI need to balance algorithmic efficiency with human oversight and contextual awareness. Organizations should leverage AI for efficiency while ensuring human decision-making maintains situational awareness and responsiveness. This balance enables them to leverage AI's strengths while mitigating associated risks.

The interaction between human intuition and machine precision is crucial for optimal performance. Bansal et al. (2019) note that successful collaboration requires both humans and AI to understand their roles and limitations, allowing them to complement each other's strengths. Teams perform best when humans have accurate mental models of AI capabilities, which in turn makes decision-making more integrated. Thus, AI should be seen as a partner that enhances human ability to navigate complex decisions.

Organizations that integrate AI with human decision-making can achieve optimal performance. Sreedharan's Human-Aware AI framework highlights the importance of AI systems that work intuitively with human users. This approach positions AI as a complement to human constraints, enabling organizations to balance machine precision with human situational awareness for optimal performance. This collaboration aligns with the principles of Strategic Minimalism, which suggests that AI enhances decision-making capabilities when it respects human rationality and context.

***Virtue Ethics and the Moral Logic of Restraint***

Virtue ethics emphasizes achieving moral excellence through moderation and self-governance (Statman, 1997). In relation to AI, the virtue of sophrosyne, which means moderation, describes being selective and thoughtful about automation technologies (Hagendorff, 2022). AlJadaan et al. (2025) argue that choosing not to automate can reflect a commitment to human dignity, fairness, and accountability in decision-making. This perspective aligns with the broader ethical considerations in AI, where decision-makers must weigh both technological capabilities and their moral consequences.

Strategic Minimalism recognizes restraint as a form of moral intelligence in AI governance. Ali et al. (2025) argue that ethical considerations should take precedence over technological advancement. They emphasize the importance of organizations understanding when to avoid automation, promoting a more thoughtful approach to ethical AI that extends beyond mere regulatory compliance and fosters a culture of ethical mindfulness regarding AI's societal impacts.

AlJadaan et al. (2025) further examine the ethical complexities of AI and automated decision-making in sectors like healthcare and finance. They assert that ethical frameworks, particularly virtue ethics, are vital for guiding AI development toward transparency and accountability, which are crucial for maintaining public trust. Thus, ethical AI governance prioritizes human oversight, ensuring that decisions reflect human values and moral responsibilities rather than being purely technology-driven.

This ethical approach to AI governance is supported by Ribeiro et al. (2025), who emphasize the link between corporate governance and AI ethics. Their article outlines how organizations can incorporate ethical principles into decision-making processes to mitigate risks associated with AI. By prioritizing ethical considerations in strategic decisions, organizations can effectively manage the relationship between technological innovation and moral responsibility.

In summary, virtue ethics provides a crucial perspective on the societal impact of AI by promoting moderation and moral restraint. By adopting Strategic Minimalism and establishing ethical AI governance frameworks, organizations can ensure their technological practices reflect essential human values, leading to a fairer and more accountable application of AI technologies.

***Frugal Innovation and Contextual Sufficiency***

Frugal innovation is crucial in resource-limited environments, focusing on effective and adaptable solutions. A study by Escudero-Cipriani et al. (2024) highlights its role in addressing global challenges, such as reducing environmental impacts and enhancing social inclusion, particularly through the adoption of cleaner technologies and responsible business practices. The research also highlights that AI can significantly enhance the scalability and efficiency of frugal solutions, supporting Hossain's (2020) assertion that it is crucial for promoting social inclusion and environmental responsibility in resource-constrained contexts.

The idea that more complexity leads to better AI performance is being challenged. According to Qin (2024), frugal AI emphasizes cost-effective solutions and sustainability without unnecessary complexities, as highlighted by Arga et al. (2025). Their research suggests that adopting a frugal approach in AI can improve performance and resource efficiency, thereby challenging traditional AI development methods. This perspective supports the need for a strategic, minimalistic approach to technology implementation in both developed and developing countries.

Strategic Minimalism emphasizes a targeted approach to AI that addresses local needs, rather than relying on complex technologies. Frimpong (2025) emphasizes the importance of aligning AI projects with the needs of resource-constrained communities. Govindan (2024) demonstrates that AI can enhance efficiency with fewer resources through sustainable frugal innovation. This strategy yields practical solutions that truly benefit communities rather than merely showcasing technology.

A minimalist approach enhances effectiveness, minimizes waste, and reduces reliance on technology. Combining frugal innovation with AI presents a significant opportunity to enhance sustainability and efficiency. By rethinking technology and innovation, stakeholders can foster environments that prioritize relevance and efficiency, promoting resource-conscious practices in the digital age.

## METHODOLOGY, THEORETICAL AND CONCEPTUAL FRAMEWORK

This paper introduces the Strategic Minimalism Framework through structured conceptual synthesis and abductive reasoning. Instead of testing pre-existing hypotheses, it combines related literature to create a concise model that outlines when a restrained approach to automation is both strategically and ethically preferable. The methodological framework advances through five iterative stages, which are summarized below.

***Step 1 – Scoping & sensitizing concepts:*** This approach applies three key concepts, namely, bounded rationality (limits on decision-making), virtue ethics (moral moderation), and frugal innovation (contextual sufficiency), to identify standard mechanisms in AI adoption, such as satisficing, moderation, and a focus on sufficiency over maximization.

***Step 2 – Construct derivation:*** From these traditions, two higher-order constructs were derived that consistently emerged across various domains.

1. Automation intensity (low → high).
2. Retained human judgment (high → low).

These map the strategic design space where AI systems and human agency develop together.

***Step 3 – Dimension and model assembly***: Crossing the two constructs creates four distinct quadrants: Virtuous Minimalism, Balanced Synergy, Automation Excess, and Performative Minimalism. The quadrant labels were selected to ensure clarity and avoid overlap between categories.

***Step 4 – Conceptual stress-testing:*** It was used to probe the model using negative cases (where "more data" fails to improve outcomes), alternative explanations (such as capability maturity or risk tiering alone), and boundary conditions (like real-time safety control with minimal human judgment). These checks ensure that the quadrants reflect real governance choices rather than just labeling artifacts.

***Step 5 – Internal coherence & parsimony:*** It assessed the framework based on standard quality criteria: internal consistency (no quadrant indicates contradictory design moves), parsimony (two axes account for various governance patterns), and explanatory reach (ability to apply to healthcare, finance, education, and public administration).

***Validation logic (conceptual):*** Without new empirical data, analytic generalization was used to assess the framework based on its ability to unify fragmented insights, generate testable propositions, and inform design and evaluation decisions in advance.

***Propositions for Empirical Falsification***

- *P1 (Proportionality Advantage).* In high-stakes situations with varying values, Virtuous Minimalism designs foster greater trust and reduce escalation events compared to Automation Excess, while maintaining the same model accuracy.
- *P2 (Sufficiency vs. Maximization*). When data quality is poor or inconsistent, systems focused on value sufficiency (such as interpretability and auditability) are more effective at reducing downstream error-correction costs than those aimed at maximization.
- *P3 (Accountability Retention).* Maintaining a minimum judgment retention ratio with clear human checkpoints reduces post-hoc liability transfer compared to designs with similar KPIs but less retained judgment.
- *P4 (Performative Risk).* Performative Minimalism leads to greater variability in results compared to Balanced Synergy.

***Evaluation & Use***

Practitioners can implement the framework through:

- design diagnostics (place current workflows in a quadrant);
- decision gates (pre-deployment checks for proportionality and judgment retention);
- governance KPIs (sufficiency metrics: interpretability, reversibility, stakeholder trust).

***Scope & Limits***

The framework offers guidance on governance design rather than specific algorithms. It is most useful for decisions involving ethical trade-offs and conflicting values but is less relevant for straightforward, low-stakes automation. The integration of these three traditions creates a unified model for AI restraint. The Strategic Minimalism Framework outlines two key dimensions:

- the level of automation (from low to high);
- the level of human judgment maintained (ranging from high to low).

Their intersection results in four key strategic placements (Figure 1):

- *Virtuous Minimalism:* the deliberate and context-aware utilization of AI to augment human insight. distinguished by proportionate automation and ethical governance.
- *Balanced Synergy:* a collaborative decision-making framework in which human and machine intelligences interact dynamically to optimize learning and performance.
- *Automation Excess:* defined by over-delegation, lack of transparency in algorithms, and weakening of accountability.
- *Performative minimalism:* symbolic restraint lacking authentic ethical reflection or strategic alignment.

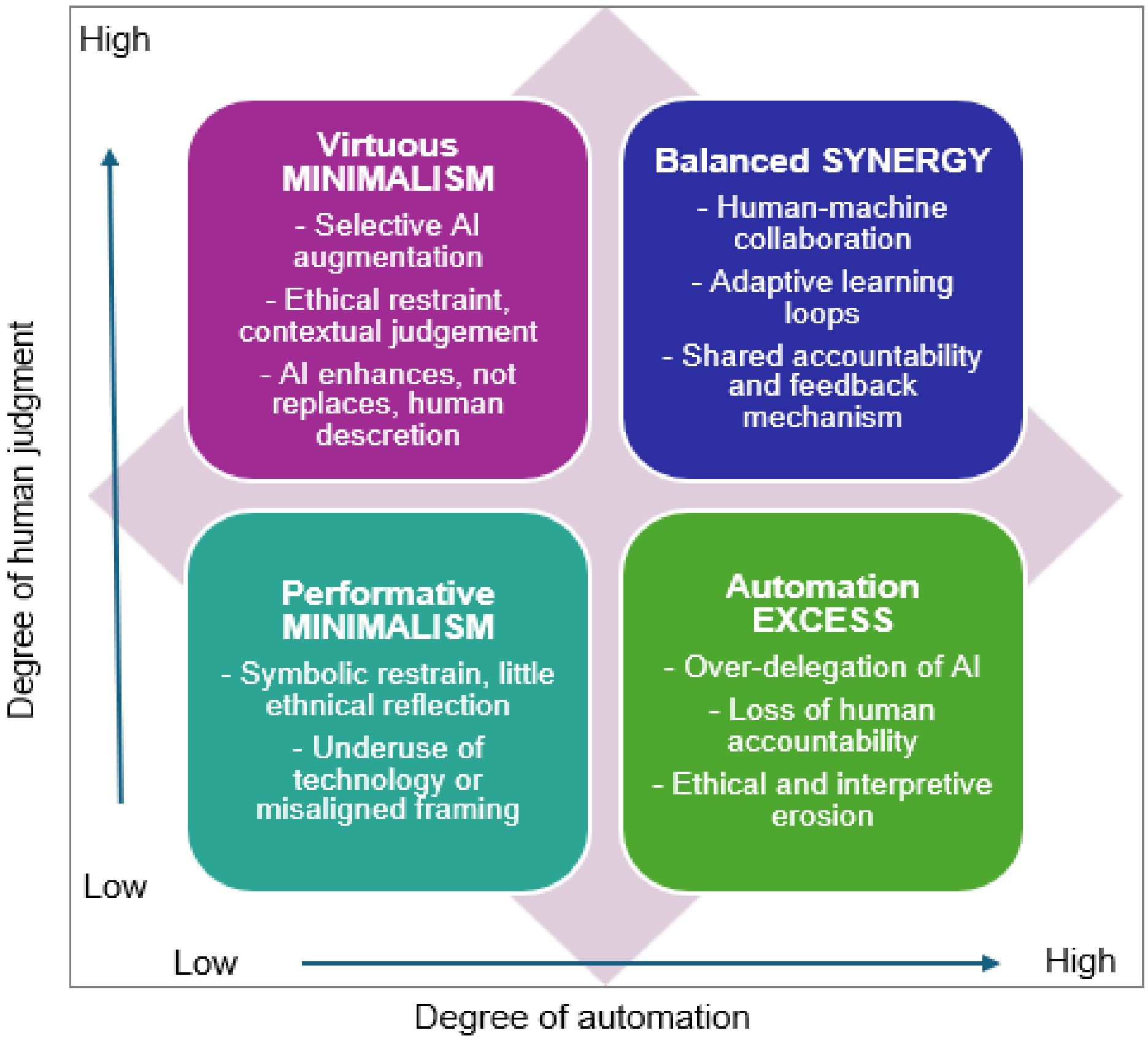

**Figure 1. The Strategic Minimalism Framework: Mapping the Balance between Automation Intensity and Retained Human Judgment**

*Source: developed by the author, 2025*

Figure 1 presents the Strategic Minimalism Framework, which categorizes organizational decisions based on the balance between automation and human judgment. The optimal quadrant, "Virtuous Minimalism", highlights effective and ethical AI use. "Automation Excess" indicates over-reliance on AI, while "Performative Minimalism" suggests only a superficial approach. "Balanced Synergy" represents a mixed decision-making environment. The model emphasizes that a true sustainable AI advantage stems from disciplined proportionality, rather than merely increasing automation.

## DISCUSSION

The Strategic Minimalism Framework provides critical guidance for leaders and policymakers seeking a balance between innovation and responsibility. While AI discussions often focus on growth, this framework emphasizes the need for selectivity and constraints as key strategic and ethical skills. It outlines four key domains that illustrate these principles.

***Managerial Implications: Leadership as Curated Intelligence***

At the managerial level, Virtuous Minimalism emphasizes that leaders should prioritize curating intelligence over simply implementing technology. They need to determine where and how much to automate, balancing efficiency with human interpretative skills. Strategic Minimalism is crucial in corporate governance and AI ethics committees, especially for organizations vulnerable to algorithmic failures and biases. This approach advocates for intentional pauses and ethical evaluations before automation decisions, framing restraint as a strategic advantage. Prioritizing human oversight alongside automation is vital for maintaining accountability and addressing the moral implications of AI.

Incorporating intentional pauses (Finlay, 2025) for ethical evaluations before starting automation processes is essential to preserve important human expertise. Additionally, building resilience requires frameworks that can handle failures and biases.

Using AI and machine learning within DevSecOps can enhance organizational resilience by proactively identifying and addressing risks (Pakalapati et al., 2023). This strategic approach enables organizations to optimize their operations while maintaining high performance, encouraging ongoing adaptation in a rapidly evolving tech landscape.

Adopting a Strategic Minimalism approach in corporate governance strengthens resilience against algorithmic failures. By prioritizing accountability, fostering ethical evaluations, and ensuring strong human oversight, organizations can navigate the complexities of AI integration while maintaining trust and legitimacy in impacted sectors.

***Ethical and Governance Implications: From Compliance to Moral Architecture***

Strategic Minimalism in corporate governance and AI ethics emphasizes proactive moral architecture rather than just reactive compliance. It prioritizes using “the right data for the right purpose”, moving away from the idea of collecting “more data for more accuracy”. Maintaining critical human oversight and focusing on appropriate performance metrics is crucial, especially in finance and healthcare, due to the ethical implications of automated decisions.

Various studies highlight the ethical complexities of AI, showing that while AI capabilities are vast, their governance necessitates rigorous frameworks for accountability and transparency. For instance, AlJadaan et al. (2025) discuss the ethical challenges of automated decision-making in sensitive areas, advocating for the use of foundational ethical frameworks such as Utilitarianism and Deontological Ethics in AI development. This approach ensures that governance assesses not only the efficiency of AI but also its qualitative impact on human well-being and ethical standards (Ramya et al., 2024).

AI systems must operate efficiently while aligning with the moral values of the societies they serve, as highlighted by Saraiva (2024). This necessitates accountability in AI design and implementation, making human oversight crucial. Additionally, Strategic Minimalism emphasizes building public trust through the implementation of ethical AI practices.

Novelli et al. (2023) outline four key goals for structuring accountability in AI (compliance, reporting, oversight, and enforcement) to address ethical challenges effectively. Organizations can foster public confidence by committing to ethical practices and ensuring that human oversight remains a key element in the deployment of AI systems.

Adopting Strategic Minimalism requires a reassessment of AI governance. Organizations should focus on developing proactive ethical frameworks that strike a balance between automated decision-making and human involvement. This strategy enhances accountability and transparency while strengthening public trust in AI, particularly in key areas such as finance, healthcare, and public administration.

***Policy Implications: Regulating for Proportionate Automation***

Strategic Minimalism offers a comprehensive framework for effective AI regulation, striking a balance between promoting innovation and upholding social and ethical standards. This model enables policymakers to categorize AI systems based on their necessity and potential for excessive automation, thereby identifying areas where enhanced regulatory oversight is warranted.

This approach aligns with significant regulations, such as the EU AI Act’s risk-based framework, while also incorporating considerations of moral proportionality and contextual relevance (European Commission, 2024; UNESCO, 2023).

Additionally, developing economies stand to gain significantly from minimalist strategies that encourage the adoption of appropriate AI solutions and diminish reliance on imported technologies, thereby strengthening local capabilities (Frimpong, 2025).

***Strategic Outlook: The Discipline of Knowing When to Stop***

Strategic Minimalism in AI governance calls for a shift from merely expanding technology to using it wisely. The achievement of responsible AI is more contingent upon the manner of implementation than on the pace at which automation is adopted. Leaders must navigate complex ethical issues in AI use, focusing on when to act, pause, or withdraw from automation.

According to Hanandeh et al. (2025), using AI thoughtfully can improve strategic planning and efficiency while reducing the risks of biases and failures (Hanandeh et al., 2025). Understanding the right timing for AI deployment can give organizations a strategic edge, enabling them to remain accountable and innovative while avoiding over-reliance on automation.

El-Gazar et al. (2025) highlight the importance of building resilience in organizational cultures that adopt AI technologies. They emphasize that leaders should prioritize ethical considerations in decision-making. Their research demonstrates that ambidextrous leadership, balancing the exploration and exploitation of technology, is essential for cultivating positive attitudes toward AI among employees. This suggests that leaders need to be discerning as they navigate technological changes. Additionally, Wang and Dai (2025) examine how the integration of AI in financial management requires strategic decision-making frameworks, advocating for careful consideration rather than rushed automation.

Strategic Minimalism redefines leadership and policy competence as the ability to know when to act, when to pause, and when to hold back. This discipline of knowing when to stop (an age-old virtue applied to modern technology) may define intelligent governance in the age of AI. Together, these domains operationalize the four quadrants of the Strategic Minimalism Framework, translating philosophical restraint into concrete governance, leadership, and policy practices.

***Policy and Practice Recommendations***

The Strategic Minimalism Framework provides a straightforward approach to applying ethical restraint in governance, management, and regulation. It helps leaders assess their organization's stance on automation and human judgment across four areas: Virtuous Minimalism, Balanced Synergy, Automation Excess, and Performative Minimalism. To ensure theoretical coherence, each recommendation is grounded in one or more of three core traditions (bounded rationality, virtue ethics, and frugal innovation) and is linked to specific quadrants of the framework.

***Institutionalize the Principle of Proportionate Automation***

Organizations should establish clear decision protocols to evaluate automation based on criteria of necessity, contextual justification, and ethical adequacy before implementation. This approach acknowledges the cognitive and contextual limits of both humans and AI. Proportional automation is not against technology; instead, it advocates for “good enough” solutions, avoiding excessive optimization that can undermine accountability. This decision-making process enables organizations to achieve a Balanced Synergy, where human and machine judgments complement each other.

***Retain a Minimum Threshold of Human Judgment in Critical Systems***

Implementing a “judgment retention ratio” is crucial for integrating moral reasoning, empathy, and interpretive discretion into high-stakes decision-making systems, such as healthcare, finance, and justice. This idea is based on virtue ethics, particularly the principle of moral moderation (sophrosyne). By retaining human judgment, organizations can adopt Virtuous Minimalism, acting not solely due to automation but because ethical considerations require it. This approach enhances accountability and preserves moral agency in automated settings.

***Embed Ethical Checkpoints in the AI Lifecycle***

Integrate ethical checkpoints at every stage of the AI lifecycle, including design, testing, deployment, and post-use evaluation. These checkpoints serve as necessary constraints, helping to ground decision-making and promoting responsible governance. This approach reduces the risk of Performative Minimalism, where ethics are only superficially acknowledged. Embedding these checkpoints ensures that moral reasoning remains ongoing, iterative, and informed by real-world data.

***Reorient Performance Metrics Toward Value Sufficiency, Not Maximization***

AI performance evaluation should focus on sufficiency, resilience, and interpretability rather than just speed or scale. Emphasizing value sufficiency promotes ethical and practical adequacy, aligning with the principles of Virtuous Minimalism and Balanced Synergy. This shift encourages organizations to prioritize wise actions over sheer output, fostering greater maturity in operations.

***Promote Contextual and Frugal AI Innovation***

In resource-constrained or developing economies, AI policy should promote innovation that is tailored to local contexts and resources. This approach builds on the frugal innovation tradition by rejecting the notion that complexity equals progress. By focusing on what is contextually sufficient, we can decrease reliance on imported technologies and enhance local capabilities. This strategy helps organizations shift from excessive automation to a focus on simplicity, adaptability, and ethical responsibility.

***Develop Regulatory Frameworks for Proportionate Automation and Accountability***

Regulators should move beyond simple "high-risk" and "low-risk" classifications and adopt a more nuanced approach that considers context, moral intensity, and human judgment. This method incorporates virtue ethics by integrating moral reasoning into legal frameworks and recognizes the limitations of predictive models. A proportional regulatory model encourages restraint, helping organizations stay within the Virtuous Minimalism and Balanced Synergy quadrants while avoiding the dangers of over-automation.

***Integrative Summary***

Theory-to-Practice Linkage: These six recommendations translate the theoretical principles of Strategic Minimalism into practical governance mechanisms.

- Bounded Rationality informs decision proportionality and context sensitivity (Recommendations 1, 3, 6).
- Virtue Ethics guides moral judgment retention and accountability preservation (Recommendations 2, 3, 6).
- Frugal Innovation anchors sufficiency and contextual adaptation (Recommendations 4, 5).

The paper presents a model of AI restraint by translating the four quadrants of the Strategic Minimalism Framework into a cycle of ethical and strategic learning. It progresses from excess to balance to virtue, illustrating how theoretical ethics can be applied in organizational policy and practice.

## CONCLUSIONS

The push for automation has made restraint an essential but often overlooked form of intelligence in the era of AI. This paper asserts that not everything that can be automated should be, and real progress in AI adoption relies on ethical decision-making. It introduces the Strategic Minimalism Framework, which incorporates bounded rationality, virtue ethics, and frugal innovation, viewing restraint as a moral and strategic adjustment rather than resistance to technology.

The framework presents two key aspects: automation intensity and retained human judgment, resulting in four organizational approaches: Virtuous Minimalism, Balanced Synergy, Automation Excess, and Performative Minimalism. These approaches demonstrate how the balance between human judgment and automation influences AI's role in either enhancing or undermining human governance.

The paper offers six policy and practice recommendations that put these ideas into action. They illustrate how bounded rationality can shape decision-making, how virtue ethics supports moral accountability, and how frugal innovation promotes adaptation. By adopting Strategic Minimalism, organizations can shift from Automation Excess to Virtuous Minimalism, building resilience, trust, and moral integrity along the way.

Conceptually, this paper contributes to Responsible AI by establishing restraint as an important, measurable virtue in intelligent governance. It demonstrates how combining theories and employing abductive reasoning can lead to the creation of models that are both normative and practical. It provides a roadmap for organizations and regulators to balance technological advancements with ethical considerations. In a world focused on automation, the key to successful intelligent organizations will be their ability to understand when and why to limit automation, reframing these limits as an essential strategy.

**Author Contributions**

Conceptualization: V. F.; data curation: V. F.; formal analysis: V. F.; funding acquisition: V. F.; investigation V. F.; methodology: V. F.; project administration: V. F.; resources: V. F.; software: V. F.; supervision: V. F.; validation: V. F.; visualization: V. F.; writing – original draft: V. F.; writing – review & editing: V. F.

**Conflicts of Interest**

The author declares no conflict of interest.

**Data Availability Statement**
Not applicable.

**Informed Consent Statement**
Not applicable.